\documentclass[prl,nofootinbib,twocolumn,superscriptaddress]{revtex4}
\def\mysection#1{{\bf #1.} }
\def\mysections#1{{\bf #1.} }
\usepackage{amssymb}
\usepackage{amsmath}
\usepackage[dvips]{graphicx}
\usepackage{longtable}
\usepackage{verbatim}
\usepackage{amsfonts}

\newcommand{\be}{\begin{equation}}
\newcommand{\ee}{\end{equation}}
\newcommand{\bea}{\begin{eqnarray}}
\newcommand{\eea}{\end{eqnarray}}
\newcommand{\beq}{\begin{equation}}
\newcommand{\eeq}{\end{equation}}
\def\beqa{\begin{eqnarray}}
\def\eeqa{\end{eqnarray}}
\newcommand{\no}{\nonumber}
\def\lsim{\mathrel{\rlap{\lower4pt\hbox{\hskip1pt$\sim$}}
    \raise1pt\hbox{$<$}}}         
\def\gsim{\mathrel{\rlap{\lower4pt\hbox{\hskip1pt$\sim$}}
    \raise1pt\hbox{$>$}}}         

\def\Bs{{\overline{B}}_s}
\begin{document}


\vspace*{-30mm}

\title{\boldmath Constraining the Phase of $B_s-\Bs$ Mixing}

\author{Yuval Grossman}\email{yuvalg@physics.technion.ac.il}
\affiliation{Department of Physics, Technion-Israel
  Institute of Technology, Technion City, Haifa 32000,
  Israel}

\author{Yosef Nir}\email{yosef.nir@weizmann.ac.il}
\affiliation{Department of Particle Physics,
  Weizmann Institute of Science, Rehovot 76100, Israel}

\author{Guy Raz}\email{guy.raz@weizmann.ac.il}
\affiliation{Department of Particle Physics,
  Weizmann Institute of Science, Rehovot 76100, Israel}

\vspace*{1cm}

\begin{abstract}
New physics contributions to $B_s-\Bs$ mixing can be parametrized by
the size ($r_s^2$) and the phase ($2\theta_s$) of the total mixing
amplitude relative to the Standard Model amplitude. The phase has so
far been unconstrained. We first use the D$\emptyset$ measurement of
the semileptonic CP asymmetry $A_{\rm SL}$ to obtain the first
constraint on the semileptonic CP asymmetry in $B_s$ decays, $A_{\rm
  SL}^s=-0.008\pm0.011$. Then we combine recent measurements by
the CDF and D$\emptyset$ collaborations --  the mass difference ($\Delta
M_s$), the width difference ($\Delta\Gamma_s$) and $A_{\rm SL}^s$ -- to
constrain $2\theta_s$. The errors on $\Delta\Gamma_s$ and $A_{\rm
  SL}^s$ should still be reduced to have a
sensitive probe of the phase, yet the central values are such that
the regions around $2\theta_s\sim3\pi/2$ and, in particular,
$2\theta_s\sim\pi/2$, are disfavored.
\end{abstract}

\maketitle

\mysection{Introduction}
Flavor changing $b\to s$ transitions are a particularly sensitive
probe of new physics. Among these, $B_s-\Bs$ mixing occupies a special
place. New physics contributions to the mixing amplitude $M_{12}^s$
can be parametrized in the most general way as follows:
\beq\label{defrthe}
M_{12}^s=r_s^2\ e^{2i\theta_s}\ (M_{12}^s)^{\rm SM},
\eeq
where $(M_{12}^s)^{\rm SM}$ is the Standard Model (SM) contribution to
the mixing amplitude. Values of $r_s^2\neq1$ and/or $2\theta_s\neq0$
would signal new physics. Assuming that the new physics can affect any
loop processes but is negligible for tree level processes, and that
the $3\times3$ CKM matrix is unitary ({\it i.e.} no quarks beyond the
known three generations), we can use various experimental measurements
to constrain the new physics parameters $r_s^2$ and $2\theta_s$:
\begin{enumerate}
\item The mass difference between the neutral $B_s$ states:
  \beq\label{dmsnp}
  \Delta M_s=(\Delta M_s)^{\rm SM}\ r_s^2.
  \eeq
\item The width difference between the neutral $B_s$ states
  \cite{Grossman:1996er,Dunietz:2000cr}:  
  \beq\label{dgsnp}
  \Delta\Gamma_s^{\rm CP}=\Delta\Gamma_s\cos2\theta_s
  =(\Delta\Gamma_s)^{\rm SM}\cos^22\theta_s.
  \eeq
\item The semileptonic asymmetry in $B_s$ decays:
  \beq\label{aslnp}
  A_{\rm SL}^s=-{\cal R}e\left(\frac{\Gamma_{12}^s}{M_{12}^s}\right)^{\rm
    SM}\frac{\sin2\theta_s}{r_s^2}. 
  \eeq
  \item The CP asymmetry in $B_s$ decays into final CP eigenstates
    such as $\psi\phi$:  
    \beq\label{sphiphi}
    S_{\psi\phi(CP=+)}=-\sin2\theta_s.
    \eeq
  \end{enumerate}
Our convention here is defined by $\Delta M_s\equiv
M_{sH}-M_{sL}$ and $\Delta\Gamma_s\equiv\Gamma_{sH}-\Gamma_{sL}$. The
observable $\Delta\Gamma_s^{\rm CP}$ is defined by
$\Delta\Gamma_s^{\rm CP}\equiv\Gamma_--\Gamma_+$, where $\Gamma_-(\Gamma_+)$
is deduced from fitting the decay rate into a final CP-odd (-even)
state assuming that it is described by a single exponential. This
assumption introduces an error of ${\cal O}(y_s^2)=0.01$
[$y_s\equiv\Delta\Gamma_s/(2\Gamma_s)$]. In
the expressions for $\Delta\Gamma_s$ and $S_{\psi\phi}$ we neglect
terms of ${\cal O}(\sin2\beta_s)=0.04$ (where 
$\beta_s=\arg[-(V_{ts}V_{tb}^*)/(V_{cs}V_{cb}^*)]$), while the
approximation for $A_{\rm SL}^s$ is good to ${\cal
  O}[(m_c^2/m_b^2)\sin2\beta_s]=0.004$. 

Until very recently, experiments gave only a lower bound on $\Delta
M_s$, a large error on $\Delta\Gamma_s$, and no meaningful information
on the CP asymmetries. Under these circumstances, there has been only
a lower bound on $r_s^2$ and no constraint at all on $2\theta_s$.

Recently, three important experimental developments took place in this
context:
\begin{itemize}
\item The CDF collaboration measured $\Delta M_s$ \cite{CDF}:
  \beq\label{dmsexp}
  \Delta M_s=17.33^{+0.42}_{-0.21}\pm0.07\ {\rm ps}^{-1}.
  \eeq
  (The D$\emptyset$ collaboration provided a milder two-sided bound
  \cite{Abazov:2006dm}.)
\item The D$\emptyset$ collaboration measured \cite{dzerodgs}
$\Delta\Gamma_s^{\rm CP}=-0.15\pm0.10^{+0.03}_{-0.04}\ {\rm
  ps}^{-1}$. Averaging this result with the earlier measurements by
CDF \cite{Acosta:2004gt} and ALEPH \cite{Barate:2000kd}, we obtain
    \beq\label{dgsexp}
    \Delta\Gamma_s^{\rm CP}=-0.22\pm0.08\ {\rm ps}^{-1}.
    \eeq
  \item The D$\emptyset$ collaboration searched for the semileptonic
    CP asymmetry \cite{dzero,dzpri}:
    \beq\label{aslexp}
    A_{\rm SL}=-0.0026\pm0.0024\pm0.0017.
    \eeq
  \end{itemize}

As obvious from eq. (\ref{dmsnp}), the main implication for new
physics of the new result for $\Delta M_s$, eq. (\ref{dmsexp}), is a
range for $r_s^2$ which can be further translated into constraints on
parameters of specific models
\cite{Ligeti:2006pm,Blanke:2006ig,Ciuchini:2006dx,Endo:2006dm,Foster:2006ze,Cheung:2006tm,Ball:2006xx}.
Here, we would like to focus instead on the phase of the mixing
amplitude $2\theta_s$. In order that a measurement of
$\Delta\Gamma_s^{\rm CP}$ can be used to constrain $\cos^22\theta_s$,
the experimental error should be at or below the level of
$(\Delta\Gamma_s)^{\rm SM}$. The new D$\emptyset$ measurement of
$\Delta\Gamma_s^{\rm CP}$ is the first to reach the required level. There are
three necessary conditions in order that a measurement of $A_{\rm SL}$ 
can be used to constrain $2\theta_s$: 
\begin{enumerate}
  \item The experimental error on $A_{\rm SL}$ should be at or below
    the level of $|\Gamma_{12}^s/M_{12}^s|^{\rm SM}$;
    \item An upper bound on $r_s^2$ should be available;
    \item An independent upper bound on $A_{\rm SL}^d$ (the
      semileptonic asymmetry in $B_d$ decays) should be available.
    \end{enumerate}
Both the D$\emptyset$ measurement of $A_{\rm SL}$ and the CDF
measurement of $\Delta M_s$ are thus crucial for our purposes, because
they satisfy, for the first time, the first and second condition,
respectively.

\mysection{Relating $A_{\rm SL}$ to $A_{\rm SL}^s$}
The semileptonic asymmetry measured at the TeVatron,
\beqa\label{defasl}
A_{\rm SL}&\equiv&\frac{\Gamma(b\bar b\to\mu^+\mu^+ X)-\Gamma(b\bar
  b\to\mu^-\mu^- X)}{\Gamma(b\bar b\to\mu^+\mu^+ X)+\Gamma(b\bar
  b\to\mu^-\mu^- X)}\no\\
&=&\frac{\Gamma_{\rm RS}^+\Gamma_{\rm WS}^+ -
  \Gamma_{\rm RS}^-\Gamma_{\rm WS}^-}{\Gamma_{\rm RS}^+\Gamma_{\rm WS}^+ +
  \Gamma_{\rm RS}^-\Gamma_{\rm WS}^-},
\eeqa
sums over all $B$-hadron decays. Given that the quark subprocesses are
$b\to \mu^-X$ and $\bar b\to\mu^+ X$, the right-sign (RS) and
wrong-sign (WS) rates can be decomposed as follows:
\beqa\label{rsws}
\Gamma_{\rm RS}^-&=&
 f_dT(\overline{B}_d\to \overline{B}_d){\overline\Gamma}_{\rm SL}^d
 +f_sT(\overline{B}_s\to \overline{B}_s){\overline\Gamma}_{\rm SL}^s
 +f_u{\overline\Gamma}_{\rm SL}^u,\no\\
\Gamma_{\rm RS}^+&=&
 f_dT({B}_d\to{B}_d)\Gamma_{\rm SL}^d
 +f_sT({B}_s\to{B}_s)\Gamma_{\rm SL}^s
 +f_u\Gamma_{\rm SL}^u,\no\\
\Gamma_{\rm WS}^-&=&
 f_dT({B}_d\to \overline{B}_d){\overline\Gamma}_{\rm SL}^d
 +f_sT({B}_s\to \overline{B}_s){\overline\Gamma}_{\rm SL}^s,\no\\
\Gamma_{\rm WS}^+&=&
 f_dT(\overline{B}_d\to {B}_d)\Gamma_{\rm SL}^d
 +f_sT(\overline{B}_s\to {B}_s)\Gamma_{\rm SL}^s,
 \eeqa
Here, $f_{q}$ is the production fraction of $B_{q}$ (we assume that
there is no production asymmetry, $f_q={\overline f}_q$), $T$ is the time
integrated probability, and $\Gamma_{\rm SL}^q$ (${\overline\Gamma}_{\rm
  SL}^q$) is the semileptonic decay rate of
$B_q$-($\overline{B}_q$-)mesons. (One should think of the $q=u$ terms
as representing all $b$-hadrons that do not mix, that is, the charged
$B$ mesons and the $\Lambda_b$ baryons.)

Within our assumptions, there is no direct CP violation in
semileptonic decays, that is, $\Gamma_{\rm SL}^q=
{\overline\Gamma}_{\rm SL}^q$. The time integrated probabilities fulfill 
$T(B_{d,s}\to B_{d,s})=T(\overline{B}_{d,s}\to\overline{B}_{d,s})$.
Consequently, we have $\Gamma_{\rm RS}^-=\Gamma_{\rm RS}^+$.
This leads to a considerable simplification of eq. (\ref{defasl}):
\beq\label{aslws}
A_{\rm SL}=\frac{\Gamma_{\rm WS}^+ - \Gamma_{\rm WS}^-}{\Gamma_{\rm WS}^+ +
  \Gamma_{\rm WS}^-}.
\eeq
Thus, the semileptonic asymmetry depends only on the wrong sign
rates. In particular, it is independent of the $B^\pm$ (and similarly
of the $\Lambda_b$) decay rates.

To a very good approximation we expect $\Gamma_{\rm SL}^d=\Gamma_{\rm
  SL}^s$ (this SU(3)-flavor equality is violated only by terms of
  ${\cal O}(m_s\Lambda_{\rm QCD}/m_b^2)$) which leads to
\beq\label{aslwssut}
A_{\rm SL}=\frac{f_dT_d^-+f_s T_s^-}{f_d T_d^++f_sT_s^+},
\eeq
where
\beq
T_q^\pm=T(\overline{B}_q\to {B}_q)\pm T({B}_q\to
\overline{B}_q).
\eeq
The relevant time integrated transition probabilities are as follows
\cite{Branco:1999fs}:
\beqa\label{ttbar}
T(B_q\to\overline{B}_q)&=&\left(\frac{1-\delta_q}{1+\delta_q}\right)\frac{Z_q}{2\Gamma_q},\no\\
T(\overline{B}_q\to{B}_q)&=&\left(\frac{1+\delta_q}{1-\delta_q}\right)\frac{Z_q}{2\Gamma_q}
\eeqa
where ($y_q=\Delta\Gamma_q/(2\Gamma_q)$, $x_q=\Delta M_q/\Gamma_q$)
\beq\label{defzq}
Z_q\equiv \frac{1}{1-y_q^2}-\frac{1}{1+x_q^2}.
\eeq
The quantity $\delta_q$ characterizes CP violation in mixing
[$\delta_q\equiv(1-|q/p|_q^2)/(1+|q/p|_q^2)$]. Given that it is small,
one can write to leading order $\delta_q= A_{\rm SL}^q/2$,
$T^-_q=A_{\rm SL}^q Z_q/\Gamma_q$ and $T^+_q=Z_q/\Gamma_q$. Taking
again the SU(3) limit, $\Gamma_d=\Gamma_s$ (the equality is violated
at high order in $1/m_b$; experimentally \cite{Barberio:2006bi}
$\tau_s/\tau_d\sim0.96\pm0.04$), we obtain \cite{diffpdg}
\beq\label{aslsut1}
A_{\rm SL}=\frac{f_dZ_dA_{\rm SL}^d+f_sZ_sA_{\rm SL}^s}{f_d Z_d+f_sZ_s}.
\eeq

Given the experimental ranges \cite{PDG} $|y_d|=0.004\pm0.019$ and 
$|y_s|=0.16\pm0.06$ we can safely neglect $y_d^2$ and $y_s^2$. (Within
our framework, we expect \cite{Beneke:2003az,Ciuchini:2003ww}\
$y_s^2\sim0.01$.) Using the experimental values \cite{Barberio:2006bi}
$f_d=0.4$, $f_s=0.1$, $x_d=0.78$ and $x_s=25.3$, we obtain 
\beq\label{aslsut}
A_{\rm SL}\simeq0.6A_{\rm SL}^d+0.4A_{\rm SL}^s.
\eeq

There are two sets of measurements that, in combination, allow us to
extract a range for $A_{\rm SL}^s$. First, we have the D$\emptyset$
measurement of $A_{\rm SL}$ (eq. (\ref{aslexp})), which we can average
together with previous measurements by the LEP experiments OPAL
\cite{Abbiendi:1998av} and ALEPH \cite{Barate:2000uk} (we neglect here
the small difference between LEP and the TeVatron regarding the
measured values of $f_{d,s}$). We find 
\beq
A_{\rm SL}=-0.0027\pm0.0029.
\eeq
Second, we have measurements of $A_{\rm SL}^d$ at the
$\Upsilon(4S)$-energy by Babar \cite{Aubert:2006nf}, Belle
\cite{Nakano:2005jb} and CLEO \cite{Jaffe:2001hz}. We find
\beq
A_{\rm SL}^d=+0.0011\pm0.0055.
\eeq

Thus, we obtain
\beq\label{aslnum}
A_{\rm SL}^s=-0.008\pm0.011.
\eeq
(One could include also the Babar measurement from hadronic modes
\cite{Aubert:2003hd}. While this is not, strictly speaking, a
measurement of $A_{\rm SL}^d$, it does give $1-|q/p|_d$. This would
change the average to $A_{\rm SL}^d=-0.0004\pm0.0055$ and, consequently,
$A_{\rm SL}^s=-0.006\pm0.011$. Our conclusions would remain
unchanged.)   

\mysection{Constraining $2\theta_s$}
Our constraints on $2\theta_s$ involve eqs. (\ref{dgsnp}) and
(\ref{aslnp}). As concerns $(\Gamma_{12}/M_{12})^{\rm SM}$, we use
\cite{Beneke:2003az} (see also \cite{Ciuchini:2003ww} for a different
calculation with similar results) 
\beq\label{gmnum}
{\cal R}e\left(\frac{\Gamma_{12}}{M_{12}}\right)^{\rm
  SM}=-0.0040\pm0.0016.
\eeq
As concerns 
$(\Delta M_s)^{\rm SM}$, we use \cite{Blanke:2006ig}
\beqa\label{dmsnum}
(\Delta M_s)^{\rm SM}&=&\frac{G_F^2}{6\pi^2}\eta_B m_{B_s}\hat
B_{B_s}F_{B_s}^2S(x_t)|V_{tb}V_{ts}|^2\no\\
&=&17.8\pm4.8\ {\rm ps}^{-1}.
\eeqa
It is important to note that the range for $|V_{ts}V_{tb}|$ is derived
using tree level  processes and CKM unitarity.
The combination of (\ref{gmnum}) and (\ref{dmsnum}) gives
\beq\label{dgsnum}
(\Delta\Gamma_s)^{\rm SM}=-0.07\pm0.03\ {\rm ps}^{-1}.
\eeq
We can now fit the new physics parameters $r_s^2$ and $2\theta_s$ to
the experimental values of eqs. (\ref{dmsexp}), (\ref{dgsexp}) and
(\ref{aslnum}) via eqs. (\ref{dmsnp}), (\ref{dgsnp}) and
(\ref{aslnp}). To do so, we use the SM estimates of
eqs. (\ref{gmnum}), (\ref{dmsnum}) and (\ref{dgsnum}).

\begin{figure}[bt]
  \centering
  \includegraphics[scale=0.6]{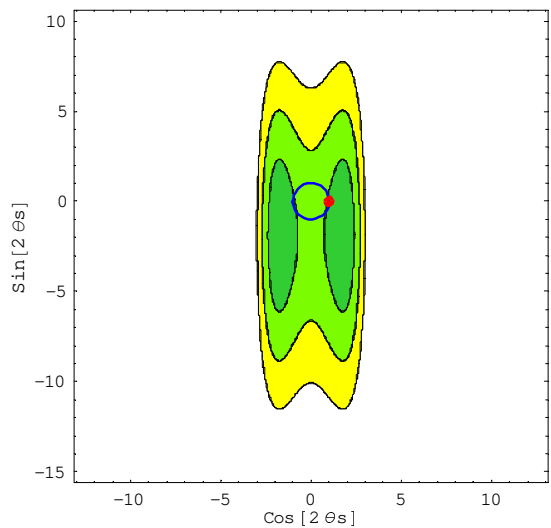}
  \caption{The constraints in the $\cos2\theta_s-\sin2\theta_s$ plane
  allowing for new physics in all loop processes. The dark green,
  light green 
  and yellow regions correspond to probability higher than 0.32,
  0.046, and 0.0027, respectively. The physical region
  ($\cos^22\theta_s+\sin^22\theta_s=1$) is along the blue circle. The SM
  point, $\cos2\theta_s=+1$, $\sin2\theta_s=0$, is marked with red.}
  \label{fig:cs}
\end{figure}

\begin{figure}[bt]
  \centering
  \includegraphics[scale=0.6]{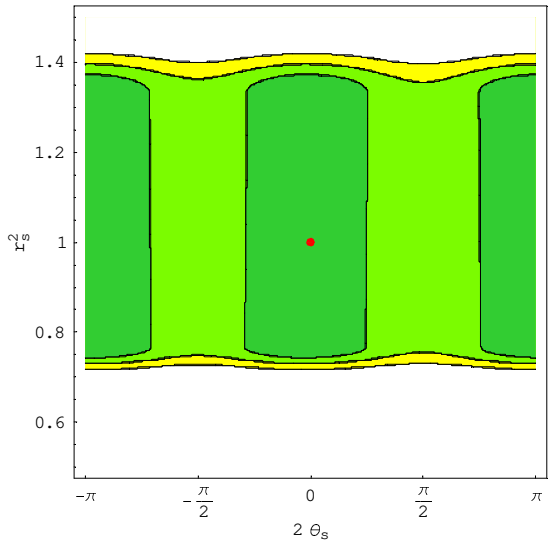}
  \caption{The constraints in the $r_s^2-2\theta_s$ plane allowing for
  new physics in all loop processes. The dark green, light green and
  yellow regions correspond to probability higher than 0.32, 0.046, and
  0.0027, respectively. The SM
  point, $2\theta_s=0$, $r^2_s=1$, is marked with red.}
  \label{fig:rsts}
\end{figure}

It is easy to understand the constraint on $r_s^2$ by simply using
eq. (\ref{dmsnp}):
\beq\label{rsfit}
r_s^2=\frac{(\Delta M_s)^{\rm exp}}{(\Delta M_s)^{\rm
    SM}}=0.97\pm0.26,
\eeq
To get a feeling for the situation concerning $2\theta_s$, we first
use eqs. (\ref{dgsnp}) and (\ref{aslnp}) separately. The
$\Delta\Gamma_s$ measurement gives 
\beq\label{ctsfit}
\cos^22\theta_s=\frac{(\Delta\Gamma_s)^{\rm CP}}{(\Delta\Gamma_s)^{\rm
    SM}}=3.1\pm1.7.
\eeq
This range disfavors (at the $1.8\sigma$ level) small
$\cos^22\theta_s$ values, that is $2\theta_s\sim\pi/2,3\pi/2$. The
$A_{\rm SL}^s$ measurement gives  
\beq\label{tsfit}
\sin2\theta_s=-\frac{A_{\rm SL}^s}{{\cal
    R}e(\Gamma_{12}^s/M_{12}^s)^{\rm SM}}
\frac{(\Delta M_s)^{\rm exp}}{(\Delta M_s)^{\rm SM}}=-1.9\pm2.8.
\eeq
This range disfavors large positive $\sin2\theta_s$ values, that is
$2\theta_s\sim\pi/2$. The combination of the two sources of
constraints should therefore disfavor the regions around
$2\theta_s\sim\pi/2,3\pi/2$, with stronger significance for the first.
This can be seen in Fig. \ref{fig:cs}, where we present the
constraints in the $\cos2\theta_s-\sin2\theta_s$ plane. In
Fig. \ref{fig:rsts} we present the constraints in the $r_s^2-2\theta_s$
plane. Note that eqs. (\ref{ctsfit}) and (\ref{tsfit}) and
Fig. \ref{fig:cs} do not take into account the correlations between
the contributions to the various observables, since they are meant to
emphasize the impact of each measurement separately. The correlations
are, however, fully taken into account in Fig. \ref{fig:rsts}.

We note that the ${\cal O}(30\%)$ error on $r_s^2$ is mainly
{\it theoretical}: it reflects the theoretical uncertainty in $(\Delta
M_s)^{\rm SM}$. In contrast, the ${\cal O}(100\%)$ error on
$\sin2\theta_s$ is mainly {\it experimental}: it comes from the error
in the determination of $A_{\rm SL}^s$. The ${\cal O}(50\%)$ error on
$\cos^22\theta_s$ has both experimental and theoretical aspects.

We learn that the constraints on $2\theta_s$ are still rather weak. In
principle, the error on $A_{\rm SL}^s$ is still a factor of three larger
than what is needed to have sensitivity to $\sin2\theta_s$. However,
since the central value for $\sin2\theta_s$ happens -- presumably due
to statistical fluctuations -- to lie below the physical region, large
positive values of $\sin2\theta_s$ are disfavored (at the $1\sigma$
level). The error on $\Delta\Gamma_s^{\rm CP}$ is closer to what is
needed to be sensitive to $2\theta_s$ and, indeed, the resulting
constraint is more significant.

We also consider a subclass of our framework, where new physics
contributions are significant only in $b\to s$ transitions. This
modifies the analysis in three ways: 
\begin{enumerate}
\item We can now extract a narrower range for $(\Delta M_s)^{\rm
 SM}$ by using, in addition to the direct calculation of
eq. (\ref{dmsnum}), an indirect calculation
\cite{ckmfitter,Bona:2005eu} that makes use of experimental
measurements of $b\to d$ (and \mbox{$s\to d$}) processes and, in
particular, identify $\Delta M_d^{\rm exp}=\Delta M_d^{\rm SM}$:
$(\Delta M_s)^{\rm SM}=21.7^{+5.9}_{-4.2}\ {\rm ps}^{-1}$
\cite{statistics}. The direct calculation of eq. (\ref{dmsnum}) and
the indirect one quoted here are essentially independent of each
other. Therefore, we average over these two results and get
\beq\label{dmsnum2}
(\Delta M_s)^{\rm SM}=19.7\pm3.5\ {\rm ps}^{-1}.
\eeq
\item We can set $A_{\rm SL}^d=0$ and then
  \beq\label{aslnum2}
  A_{\rm SL}^s\simeq2.5A_{\rm SL}=-0.007\pm0.007.
  \eeq
\item We can now use (\ref{dmsnum2}) to obtain a more precise estimate
  of $(\Delta\Gamma_s)^{\rm SM}$:
  \beq\label{dgsnum2}
  (\Delta\Gamma_s)^{\rm SM}=-0.08\pm0.03\ {\rm ps}^{-1}.
  \eeq
\end{enumerate}
Now we get
\beq
r_s^2=0.88\pm0.16,
\eeq
\beq\label{costs2}
\cos^22\theta_s=2.8\pm1.6,
\eeq
\beq\label{sints2}
\sin2\theta_s=-1.4\pm1.6.
\eeq
The situation is then quite similar to the first scenario.
The smaller central value and smaller error on $r_s^2$ and on 
$\cos^22\theta_s$, compared to eqs. (\ref{rsfit}) and (\ref{ctsfit}),
respectively, correspond to the larger central value and smaller
theoretical error in eq. (\ref{dmsnum2}) compared to
eq. (\ref{dmsnum}). In contrast, the higher central value and smaller
error on $\sin2\theta_s$, compared to  eq. (\ref{tsfit}), are both
mainly a result of the shift in the central value of $r_s^2$ and, in
particular, little affected by the smaller error on $(\Delta
M_s)^{\rm SM}$.

\begin{figure}[bt]
  \centering
  \includegraphics[scale=0.6]{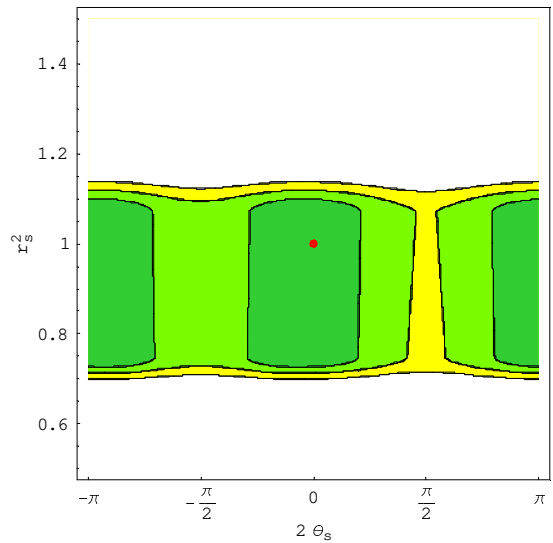}
  \caption{The constraints in the $r_s^2-2\theta_s$ plane allowing for
  new physics in $b\to s$ loop processes only. The dark green, light
  green and yellow regions correspond to probability higher than 0.32,
  0.046, and  0.0027, respectively. The SM
  point, $2\theta_s=0$, $r^2_s=1$, is marked with red.}
  \label{fig:bs}
\end{figure}

We show the contraints in the $r_s^2-2\theta_s$ plane in
Fig. \ref{fig:bs}. As can be seen in the Figure, $2\theta_s=\pi/2$ is
disfavored at the $2\sigma$ level. 

\mysection{Conclusions}
The measurement of $A_{\rm SL}$ by D$\emptyset$ probes CP violation in
$B_s-\Bs$ mixing, $A_{\rm SL}^s=-0.008\pm0.011$. 
In combination  with the measurement of $\Delta M_s$ by CDF, and the
measurements of $\Delta\Gamma_s^{\rm CP}$ by D$\emptyset$ and CDF, the CP
violating phase of the mixing amplitude is constrained for the first
time. The constraints are still weak. Since experiments favor
large values of $\Delta\Gamma_s$ compared to the SM value, small
values of $\cos^22\theta_s$ ({\it i.e.} $2\theta_s\sim\pi/2,3\pi/2$)
are disfavored. Furthermore, since experiments favor a
negative $A_{\rm SL}^s$ (see eqs. (\ref{aslnum}) and (\ref{aslnum2}))
and ${\cal R}e(\Gamma_{12}^s/M_{12}^s)^{\rm SM}$ is negative, large
positive values of $\sin2\theta_s$ ({\it i.e.} $2\theta_s\sim\pi/2$)
are disfavored even more strongly. 

To improve the constraint, smaller experimental errors on
$\Delta\Gamma_s$ and on $A_{\rm SL}$ are welcome. Note however that a
similar improvement in the measurement of $A_{\rm SL}^d$ (see
eq. (\ref{aslsut})) is also required. Thus, the accuracy in
determining $A_{\rm SL}^s$ depends on both high energy
hadron machines and $\Upsilon(4S)$-energy B factories.

In principle, $A_{\rm SL}^s$ could also be extracted from
measurements at hadron colliders only. To do this one
needs, in addition to the measurement of $A_{\rm SL}$, another
measurement of a CP asymmetry in semileptonic decays, with a different
weight of $B_d$ and $B_s$ in the sample. (For example, requiring at
least one kaon in the final state would enhance the fraction of
$B_s$.)

Of course, the phase $2\theta_s$ will be strongly constrained once
$S_{\psi\phi}$ is measured. Then the combination of the four
measuerements -- $\Delta M_s$, $\Delta\Gamma_s$, $A_{\rm SL}^s$ and
$S_{\psi\phi}$ -- will 
provide a test of the assumption that new physics affects only loop
processes \cite{Laplace:2002ik,Ligeti:2006pm,Blanke:2006ig}. The
strength of this test will, however, be limited by theoretical
uncertainties, particularly by the calculation of $\Gamma_{12}^{\rm
  SM}$.  

\mysections{Acknowledgments} We are grateful to Daria Zieminska for
drawing our attention to the relevance of $\Delta\Gamma_s$ to our
analysis and for providing us with further valuable information and
advice. We are grateful to Guennadi Borissov for clarifying to us the
way in which $A_{\rm SL}$ was determined by D0. We thank Andrzej Buras,
Andreas H\"ocker, Zoltan Ligeti and Marie-H\'el\`ene Schune for useful
discussions. This  
project was supported by the Albert Einstein Minerva Center for
Theoretical Physics, and by EEC RTN contract HPRN-CT-00292-2002. The
work of Y.G. is supported in part by the Israel Science Foundation
under Grant No. 378/05. The research of Y.N. is
supported by the Israel Science Foundation founded by the Israel
Academy of Sciences and Humanities, and by a grant
from the United States-Israel Binational Science Foundation (BSF),
Jerusalem, Israel. 


\end{document}